\documentclass{aa}  

\usepackage{graphicx}
\usepackage[varg]{txfonts}
\usepackage{pdflscape}
\begin{document}
   \title{A search for hydrogenated fullerenes in fullerene-containing planetary nebulae}

   \titlerunning{}

   \author{J. J.  D\'{\i}az-Luis\inst{1,2}, D. A.
   Garc\'{\i}a-Hern\'andez\inst{1,2}, A. Manchado\inst{1,2,3} \and F. Cataldo\inst{4,5} }
	  
\authorrunning{D\'{\i}az-Luis et al.}

   \institute{Instituto de Astrof\'{\i}sica de Canarias, C/ Via L\'actea s/n, E$-$38205 La Laguna, Spain \email{jdiaz@iac.es, agarcia@iac.es}
         \and Departamento de Astrof\'{\i}sica, Universidad de La Laguna (ULL), E$-$38206 La Laguna, Spain
         \and Consejo Superior de Investigaciones Cient\'{\i}ficas, Madrid, Spain
         \and INAF- Osservatorio Astrofisico di Catania, Via S. Sofia 78, Catania 95123, Italy
         \and Actinium Chemical Research srl, Via Casilina 1626/A, 00133 Rome, Italy
             }

   \date{Received xx, 2016; accepted xx x, 2016}

 
\abstract {Detections of C$_{60}$ and C$_{70}$ fullerenes in planetary nebulae (PNe) of
the Magellanic Clouds and of our own Galaxy have raised the idea that other forms of
carbon such as hydrogenated fullerenes (fulleranes like C$_{60}$H$_{36}$ and
C$_{60}$H$_{18}$), buckyonions, and carbon nanotubes, may be widespread in the Universe.
Here we present VLT/ISAAC spectra (R $\sim$600) in the 2.9-4.1 $\mu$m spectral region for
the Galactic PNe Tc 1 and M 1-20, which have been used to search for fullerene-based
molecules in their fullerene-rich circumstellar environments. We report the non-detection
of the most intense infrared bands of several fulleranes around $\sim$3.4-3.6
$\mu$m in both PNe. We conclude that if fulleranes are present in the
fullerene-containing circumstellar environments of these PNe, then they seem to be by far
less abundant than C$_{60}$ and C$_{70}$. Our non-detections together with the
(tentative) fulleranes detection in the proto-PN IRAS 01005$+$7910 suggest that
fulleranes may be formed in the short transition phase between AGB stars and PNe
but they are quickly destroyed by the UV radiation field from the central star.}

\keywords{Astrochemistry --- Line: identification --- circumstellar matter ---
ISM: molecules --- planetary Nebulae: individual: Tc 1, M 1-20}

\maketitle


\section{Introduction}

Fullerenes, highly resistant and stable tridimensional molecules formed exclusively by
carbon atoms, have attracted much attention since their discovery at laboratory by Kroto
et al. (1985); C$_{60}$ and C$_{70}$ are the most abundant fullerenes. Fullerenes and
fullerene-based molecules such as hydrogenated fullerenes and buckyonions may explain
some astrophysical phenomena like the intense UV  absorption band at 217 nm (e.g.,
Cataldo \& Iglesias-Groth 2009) or the so-called diffuse interstellar bands (DIBs; see
e.g., Cox 2011; Iglesias-Groth 2007; Garc\'{\i}a-Hern\'andez  \& D\'iaz-Luis 2013;
D\'iaz-Luis et al. 2015). Due to the remarkable stability of fullerenes against intense
radiation and ionization, these molecules are good candidates to be  widespread in
insterstellar/circumstellar media. Their presence in astrophysical  environments have
been recently confirmed by the {\it Spitzer Space Telescope} via the detection of the
mid-infrared C$_{60}$ and C$_{70}$ fullerene features in the spectrum of the young 
Planetary Nebula (PN) Tc 1 (Cami et al. 2010).

The detection of fullerenes in Tc 1 was interpreted as due to the H-poor
conditions in the nebular core (since there is no polycyclic aromatic
hydrocarbons (PAHs) emission; Cami et al. 2010), in agreement with some
laboratory studies (e.g., Kroto  et al. 1985), which show that fullerenes are
efficiently produced in the absence of hydrogen. However, it has been
demonstrated that fullerenes are efficiently formed in H-rich circumstellar
environments only (Garc\'{\i}a-Hern\'andez et al. 2010, 2011b), challenging our
previous understanding of the fullerenes formation in space. In particular, the
simultaneous detection of C$_{60}$ fullerenes and PAHs in several PNe with
normal hydrogen abundances has been reported (Garc\'{\i}a-Hern\'andez et al.
2010, 2011a, 2012; Otsuka et al. 2014). Garc\'{\i}a-Hern\'andez et al. (2010)
propose that both fullerenes and PAHs in H-rich circumstellar ejecta may form
from the photochemical processing of hydrogenated amorphous carbon (HAC) in
agreement with the experimental results of Scott et al. (1997)\footnote{Duley \&
Hu (2012) suggest that the IR features at $\sim$7.0, 8.5, 17.4, and 18.8 $\mu$m
detected in objects with PAH-like-dominated IR spectra, such as those of R
Coronae Borealis stars (Garc\'{\i}a-Hern\'andez et al. 2011b), reflection
nebulae (Sellgren et al. 2010), or proto-PNe (Zhang \& Kwok 2011) should be
attributed to proto-fullerenes (or fullerene precursors) instead of C$_{60}$.}.
In addition, the first evidence for the possible detection of planar
C$_{24}$ (a piece of graphene) in some of these fullerene-containing sources has
been reported (Garc\'{\i}a-Hern\'andez et al. 2011a), raising the idea that
other forms of carbon such as hydrogenated fullerenes (fulleranes such as
C$_{60}$H$_{36}$ and C$_{60}$H$_{18}$), buckyonions or carbon nanotubes, may be
widespread in the Universe. 

Duley \& Williams (2011) predict the production of fullerenes via thermal heating of HAC
and this process may explain the detection of fullerenes in H-rich circumstellar
environments (e.g., Garc\'{\i}a-Hern\'andez et al. 2010). Interestingly, they predict
also the formation of hydrogenated fullerenes (fulleranes). On the other hand,
Iglesias-Groth et al. (2012) found that the 3.44 and 3.55 $\mu$m  bands of several
fulleranes display molar extinction coefficients which are similar to those  of
the 17.4 and 18.8 $\mu$m bands of the isolated C$_{60}$ molecule (Iglesias-Groth et al.
2011).  Thus, it could be expected to find fulleranes with line intensities
similar to the ones already measured  with {\it Spitzer} at 17.4 and 18.8 $\mu$m in
fullerene-containing PNe. Indeed, Zhang \& Kwok (2013) have tentatively detected
fulleranes in the {\it Infrared Space Observatory} (ISO) spectrum of the proto-PN
IRAS 01005$+$7910, where three emission peaks at $\sim$3.48, 3.51, and 3.58 $\mu$m in the
C-H stretching region seem to be present. 

More recently, several fullerene-based molecules like fullerene/PAH adducts have
been synthesized and characterized at laboratory (e.g., Garc\'{\i}a-Hern\'andez
et al. 2013; Cataldo et al. 2014, 2015). Remarkably, fullerene/PAH adducts
such as C$_{60}$/anthracene display mid-IR ($\lambda$ $>$ 5 $\mu$m) spectral
features strikingly coincident with those from C$_{60}$ and C$_{70}$
(Garc\'{\i}a-Hern\'andez et al. 2013) and it is still no clear if such species
could contribute to the observed C$_{60}$ (and C$_{70}$) features in
fullerene-rich environments. Laboratory spectra of C$_{60}$/anthracene adducts
display the typical aromatic bands around 3.3 $\mu$m as well as aliphatic C$-$H
bands at $\sim$3.39, 3.43 and 3.52 $\mu$m, which are not present in the C$_{60}$
and C$_{70}$ spectra. In principle, this could be used to elucidate the possible
carrier (e.g., C$_{60}$ versus C$_{60}$/PAH adducts) of the mid-IR features seen
in fullerene PNe. However, the C$-$H stretching bands are intrinsically weaker
than the other emission features at longer wavelengths and the presence of these
C$-$H stretching bands (typical for CH$_{2}$ and CH$_{3}$ groups of several
carbonaceous materials) is not yet completely understood
(Garc\'{\i}a-Hern\'andez et al. 2013); which complicate the search of these
fullerene-related species in space. 

In this paper, we present VLT/ISAAC spectroscopy (the 2.9-4.1 $\mu$m spectral
region) of two PNe with fullerenes (Tc 1 and M 1-20). An overview of the
spectroscopic observations is presented in Section 2; together with a
summary of the nebular emission lines observed. Section 3 presents a brief
discussion of the detection of the 3.3 $\mu$m unidentified infrared emission
(UIE) feature in our VLT/ISAAC spectrum of M 1-20 and its possible carriers.
Section 4 discusses the non-detection of infrared emission bands (e.g.,
from fulleranes) in our spectra. The conclusions of our work are given in
Section 5.


\section{Mid-infrared VLT/ISAAC spectroscopy of PNe with fullerenes}

We acquired 3-4 $\mu$m infrared (IR) spectra of the fullerene PNe Tc 1 (W1[3.35$\mu$m]$_{WISE}$=8.19,
E(B-V)=0.23; Cutri et al. 2012; Williams et al. 2008) and M 1-20
(W1[3.35$\mu$m]$_{WISE}$=9.65, E(B-V)=0.80; Cutri et al. 2012;
Wang \& Liu 2007) with a S/N (at the continuum in the
final IR spectra) of $\sim$26 and 11, respectively. Tc 1 displays a
fullerene-dominated spectrum with no clear signs of PAHs, while M 1-20 also
shows weak PAH-like features (see, e.g., Garc\'{\i}a-Hern\'andez et al. 2010).
Table 1 lists some observational parameters such as Galactic coordinates, colour
excess and radial velocity for the two fullerene PNe in our sample and the
corresponding telluric/flux stars for each PN.

The observations of Tc 1, M 1-20, and their corresponding telluric/flux standards were
carried out at the ESO VLT (Paranal, Chile) with ISAAC in service mode between 14-23 July
2013. We used the LWS-LR/3550 nm set-up with the 0.6" slit oriented at a position angle
of 0$^\circ$. This configuration covers the spectral range 2.9-4.1 $\mu$m and
gives a resolving power of $\sim$600, which is required to cleanly separate the
3-4 $\mu$m features of several fullerene-based molecules as seen at laboratory. During
the observations, the seeing was about 0.5-0.6 and 1.5-1.7 arcsec for Tc 1 and M 1-20
(and their corresponding standards), respectively.

\begin{table*}
\tiny
\caption{\label{t1}Observational parameters of fullerene PNe and their comparison stars.}
\centering
\small\begin{tabular}{lccccclccccc}
\hline\hline
Object  & l & b &  E(B-V)  & V$_{r}$ & Ref\tablefootmark{a} & Telluric/flux star & l & b & SpT & V$_{r}$ &  Ref\tablefootmark{a}  \\
\hline
Tc 1    &  345.2375  & $-$08.8350 & 0.23 & $-$94.0 & 1, 2     & HR 7446   & 31.7709  & $-$13.2866 & B0.5III & $-$19.4  & 4, 5  \\
M 1-20  &  6.187     &     8.362  & 0.80 &    75.0 & 2, 3     & HIP 92519 & 351.7764 & $-$18.6123 & G0V     & 70.9     & 4, 6  \\
\hline\hline
\end{tabular}
\tablefoot{
\\
\tablefoottext{a}{References. (1) Williams et al. (2008); (2) Beaulieu et al. (1999); (3) Wang \& Liu (2007);
(4) De Bruijne \& Eilers (2012); (5) Wegner (2003); (6) Soubiran et al. (2013).}
}
\end{table*}

The spectra were obtained combining the chopping technique (moving the secondary
mirror of the telescope once every few seconds) with telescope nodding. We used
a total integration time of 38 min for each observing block (OB). For Tc 1, we
obtained three OBs of 26 individual exposures each, giving a total exposure time
of $\sim$1.9 hours, and for M 1-20, one OB of $\sim$38 min. Raw spectra were
processed by the ISAAC data reduction
pipeline\footnote{https://www.eso.org/sci/software/pipelines/isaac/isaac-pipe-recipes.html} 
in conjunction with the data browsing tool
GASGANO\footnote{https://www.eso.org/sci/software/gasgano.html}. In short, i)
flat-fields are combined to produce a master flat-field; ii) the wavelength
calibration and ISAAC slit curvature distortion is computed using OH sky lines;
and iii) the removal of the high degree of curvature of ISAAC spectra is done by
calculating the spectra curvature using a star moving across the slit. Thus,
science frames were reduced using the products of the pipeline calibration
recipes. The produced 2D image for each PN is then used to extract the one
dimensional spectrum across the defined apertures with IRAF\footnote{Image
Reduction and Analysis Facility (IRAF) software is distributed by the National
Optical Astronomy Observatories, which is operated by the Association of
Universities for Research in Astronomy, Inc., under cooperative agreement with
the National Science Foundation.}. We note that the continuum emission is not
extended in both PNe but in order to cover the nebular hydrogen lines in the 2D
images, we needed to define an aperture of 78 and 10 pixels for Tc 1 and M 1-20,
which translate (scale of 0.15 arcsec/pixel) into nebular extensions of
$\sim$11.5 and $\sim$1.5 arcsec, respectively. Finally, the extracted spectra
were combined to produce a final reduced science spectrum. This process was done
also for the telluric stars, which are used also for flux calibration (see
below).

We have used the telluric stars to determine the sensitivity and extinction functions to
flux calibrate our science spectra with standard tasks in IRAF. We have assumed that the
two stars behave approximately like blackbodies. There are two parameters needed to scale
the blackbody to the observation: the magnitude of the star in the same band as the
observation (L band at 3.4 $\mu$m) and the effective temperature. The estimated flux
errors are approximately 30-40 \%. Finally, telluric correction was made using the
telluric star for each PN. 


\subsection{Nebular emission lines}

In this section, we give an overview of the nebular emission lines observed in the 2.9 to
4.1 $\mu$m region. This spectral range is dominated by nebular emission lines of
hydrogen.

Our list of features in Tc 1 and M 1-20 is shown in Table 2, where we give the measured
central wavelength ($\lambda$$_{c}$), the full width at half maximum (FWHM), the
equivalent width (EQW) and the integrated flux\footnote{The feature parameters were
measured with IRAF by assuming a Gaussian profile.}.

\begin{table*}
\tiny
\caption{\label{t2}Nebular emission lines identified in the 2.9-4.1 $\mu$m spectra of M 1-20 and Tc 1.}
\centering
\small\begin{tabular}{lcccccc}
\hline\hline
        & M 1-20          &                   &                                  & Tc 1            &                    &                                  \\       
Element & $\lambda$$_{c}$ &  FWHM             & FLUX\tablefootmark{a}            & $\lambda$$_{c}$ & FWHM               & FLUX\tablefootmark{a}            \\                                                               
        &  ($\mu$m)       & (10$^{-4}$$\mu$m) & (10$^{-15}$ergcm$^{-2}$s$^{-1}$) & ($\mu$m)        & (10$^{-4}$$\mu$m)  & (10$^{-15}$ergcm$^{-2}$s$^{-1}$) \\                                                               
\hline
  H I (Pf$_{\epsilon}$)       & 3.07 & 43.68    &  4.20  &       &        &       \\
  UIE              & 3.31 & 536.10   &  17.20 &       &        &       \\
  H I (Pf$_{\delta}$) & 3.32 & 41.05 &  6.49  & 3.30  & 37.88  &  0.28 \\                            
  H I (Pf$_{\gamma}$) & 3.75 & 45.96 &  5.74  & 3.74  & 47.32  &  0.43 \\                            
  H I (6-17)       & 3.76 & 40.95    &  0.33  & 3.75  & 42.43  &  0.03 \\                           
  H I (6-15)       & 3.91 & 40.45    &  0.54  & 3.90  & 37.77  &  0.04 \\ 
  He I (4-5)       & 4.04 & 25.01    &  0.86  & 4.04  & 43.17  &  0.11 \\
  H I (Br$_{\alpha}$) & 4.05 & 47.59 &  46.70 & 4.05  & 48.48  &  3.49 \\                                    
\hline\hline
\end{tabular}
\tablefoot{
\tablefoottext{a}{Estimated flux errors are of the order of $\sim$30\%-40\%.}
}
\end{table*}

The M 1-20 spectrum includes the atomic hydrogen lines Pf$_{\epsilon}$ at
3.07 $\mu$m, a blend of the UIE feature at 3.31 with Pf$_{\delta}$ at 3.32
$\mu$m, Pf$_{\gamma}$ at 3.75 $\mu$m and Br$_{\alpha}$ at 4.05 $\mu$m. It also
includes the tentatively identified H I (6-17), H I (6-15) and He I (4-5) lines
at 3.76, 3.91 and 4.04 $\mu$m, respectively. In Figure 1, we display the M 1-20
spectrum in the 2.9-4.1 $\mu$m range. 

The Tc 1 spectrum includes the atomic hydrogen lines Pf$_{\delta}$,
Pf$_{\gamma}$, the tentatively identified H I (6-17) and H I (6-15) lines and
Br$_{\alpha}$. It also includes the tentative He I (4-5) line at 4.04 $\mu$m. In
Figure 2, we display the Tc 1 2.9-4.1 $\mu$m spectrum.

 \begin{figure*}
   \centering
   \includegraphics[angle=0,scale=.70]{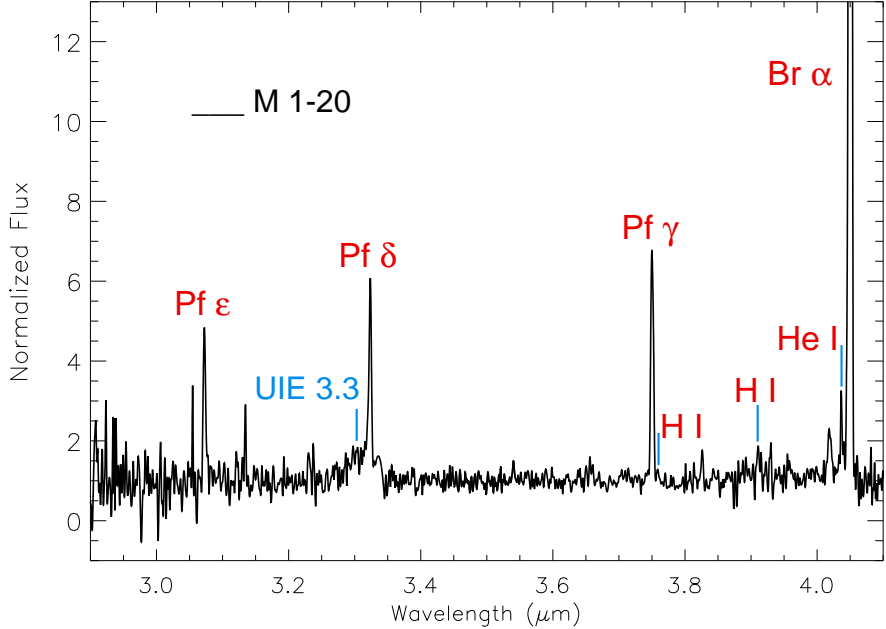}
    \caption{VLT/ISAAC spectrum of PN M 1-20. The atomic H lines of the Pfund series
    (Pf$_{\epsilon}$, Pf$_{\delta}$ and Pf$_{\gamma}$), H I (6-17), H I (6-15) and Br$_{\alpha}$
    as well as the UIE feature at 3.3 $\mu$m and the line of He I (4-5) are indicated.
 \label{Fig1}}
    \end{figure*}

 \begin{figure*}
   \centering
   \includegraphics[angle=0,scale=.70]{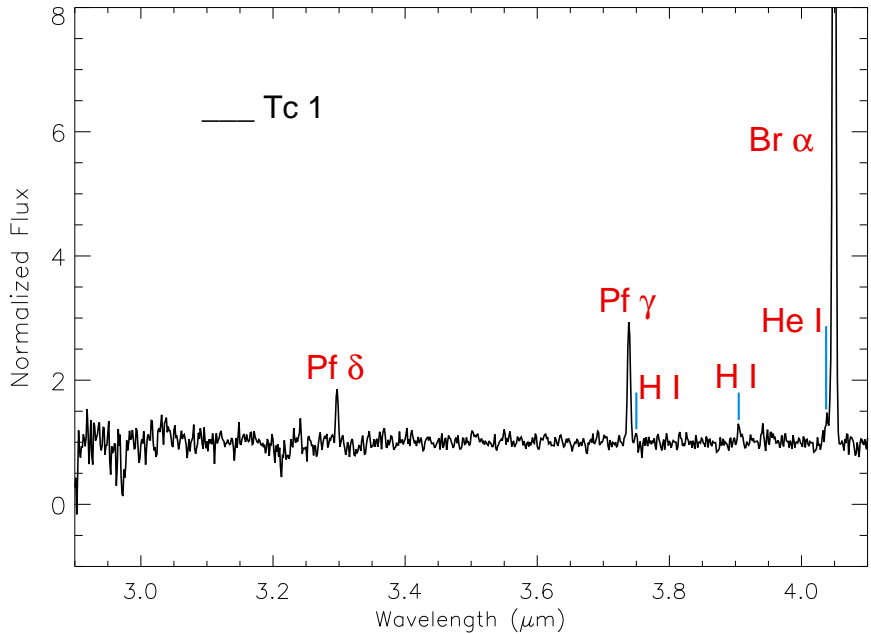}
    \caption{VLT/ISAAC spectrum of Tc 1. The atomic H lines of the Pfund series
    (Pf$_{\delta}$ and Pf$_{\gamma}$), H I (6-17), H I (6-15) and Br$_{\alpha}$,
    as well as the line of He I (4-5) are indicated.
 \label{Fig2}}
    \end{figure*}
    

\section{The 3.3 $\mu$m UIE feature}

We clearly detect the 3.3 $\mu$m UIE feature in M 1-20, while this feature is
completely lacking in Tc 1. This is consistent with the {\it Spitzer} spectra; M
1-20 displays UIE features (PAH-like) at 6.2, 7.7, 8.6 and 11.3 $\mu$m, but
these are absent in Tc 1 (e.g., Garc\'{\i}a-Hern\'andez et al. 2010). The
detected UIE feature in the M 1-20 spectrum was fitted with a Gaussian to
determine the feature's FWHM and central wavelength. Due to the emission line of
H I (Pf$_{\delta}$) at 3.32 $\mu$m, it was necessary to deblend both features.
The average central wavelength  of the UIE feature was 3.31 $\mu$m, with a FWHM
of 0.05 $\mu$m. This is consistent with the values measured by Tokunaga et al.
(1991) and Smith \& McLean (2008) (average values of $\lambda$$_{c}$ $\sim$3.29
$\mu$m and FWHM $\sim$0.04 $\mu$m) in extended objects such as PNe, (proto-)
PNe, and H II regions. In Figure 1, we can see the 3.3 $\mu$m UIE feature in the
M 1-20 spectrum blended with the Pf$_{\delta}$ hydrogen emission line  at 3.32
$\mu$m. 

\subsection{Possible carriers of the 3.3 $\mu$m UIE feature}

The 3.3 $\mu$m feature belongs to the group of UIE features or aromatic infrared
bands (AIBs) at 3.3, 6.2, 7.7, 8.6, 11.3 and 12.7 $\mu$m. These bands may be
accompanied by aliphatic bands at 3.4, 6.9 and 7.3 $\mu$m and unassigned
features at 15.8, 16.4, 17.4, 17.8 and 18.9 $\mu$m (Jourdain de Muizon et al.
1990; Kwok et al. 1999; Chiar et al. 2000; Sturm et al. 2000; Sellgren et al.
2007)\footnote{It is now known that the 17.4 and 18.9 $\mu$m features are due to
fullerenes.}, and the broad emission features at $\sim$6-9, 9-13, 15-20, and
25-35 $\mu$m, which are believed to be produced by a carbonaceous mixture of
aromatic and aliphatic structures (e.g., HAC, coal, petroleum fractions, etc.)
and/or their decomposition products (Garc\'{\i}a-Hern\'andez et al. 2010, 2012;
Kwok \& Zhang 2011). The 11.3 $\mu$m UIE feature is found to be correlated with
the 3.3 $\mu$m feature (Russell et al. 1977), suggesting a common origin for the
two features. This correlation holds for fullerene PNe; we detect the weak UIE
feature at 3.3 $\mu$m in our VLT/ISAAC spectrum and the M 1-20 {\it Spitzer}
spectrum shows a very strong aromatic-like infrared band at 11.3 $\mu$m
(Garc\'{\i}a-Hern\'andez et al. 2012), while Tc 1 displays a very weak emission
at 11.3 $\mu$m and it does not show the feature at 3.3 $\mu$m.

A wide variety of molecules have been proposed as possible carriers of these 3.3
and  11.3 $\mu$m UIE bands. Nowadays, the most accepted idea is that they
originate from the C-H vibration modes of aromatic compounds or PAHs (e.g.,
Duley \& Williams 1981; Leger \& Puget 1984; Allamandola, Tielens \& Barker
1985). However, Sadjadi et al. (2015) calculated the out-of-plane (OOP) bending
mode frequencies of 60 neutral PAH molecules and found that it is difficult to
fit the 11.3 $\mu$m UIE feature. This feature cannot be fitted by superpositions
of pure PAH molecules and O- (and or Mg-) containing species are needed to
achieve a good fit, suggesting that the PAHdb model (Bauschlicher et al. 2010;
Boersma et al. 2014) has difficulties to explain the UIE phenomenon (see Sadjadi
et al. 2015 for more details). Maltseva et al. (2015) also found difficulties
for predicting high-resolution experimental absorption spectra of PAHs in the
3-$\mu$m region with harmonic density functional theory (DFT) calculations.
Finally, Gadallah et al. (2013) also found that the 3.3 $\mu$m and 6.2
$\mu$m AIBs are also clearly observed in the spectrum of heated HAC dust. 

On the other hand, Duley \& Williams (2011) suggest a model in which the
astronomical emission at 3.3 $\mu$m can be explained by the heating of HAC dust
via the release of stored chemical energy. This energy would be sufficient to
heat dust grains (with sizes of $\sim$50-1000 \AA) to temperatures at which they
can emit at 3.3 $\mu$m. This alternative to the PAH hypothesis involves a solid
material with a mix of aliphatic and aromatic structures (i.e., HACs), which may
explain the broad emission features at $\sim$6-9, 9-13, 15-20, and 25-35 $\mu$m
detected in PNe with fullerenes (Garc\'{\i}a-Hern\'andez et al. 2010, 2012) and
most C-rich proto-PNe. Moreover, laboratory experiments show that the
products of destruction of HAC grains are PAHs and fullerenes (Scott et al.
1997), something that could explain the detection of both types of molecules in
some fullerene PNe (Garc\'{\i}a-Hern\'andez et al. 2010). Unfortunately, our
VLT/ISAAC observations do not add much information about the real carrier (e.g.,
PAHs vs. HACs) of the 3.3 $\mu$m emission. The non-detection of 3.3 $\mu$m
emission in Tc 1 may suggest a different spatial distribution of the 3.3 $\mu$m
carrier and the C$_{60}$ (and C$_{70}$) fullerenes. Bernard-Salas et al. (2012)
already found that the {\it Spitzer} C$_{60}$ 8.5 $\mu$m and weak 11.2 $\mu$m
emission (which likely share the same carrier with the 3.3 $\mu$m emission, see
above) are extended but they peak at opposite directions from the Tc 1 central
star\footnote{Sellgren et al. (2010) reported a similar separation in the
spatial distribution of the fullerene and PAH emission seen in the
fullerene-containing reflection nebula NGC 7023.}. The {\it Spitzer}
observations (slit of $\sim$4" x 57" at a P.A. of 0 degrees) show extended
emission up to $\sim$22" (Bernard-Salas et al. 2012), while our VLT/ISAAC
spectra, taken with a smaller slit of 0.6 arcsec, show no extended emission at
3.3 $\mu$m. Thus, it could be possible that we miss the weak 3.3 $\mu$m emission
in the ISAAC observations (e.g., a small column density throughout the
circumstellar envelope). In the case of the more compact (apparent size of
$\sim$2 arcsec) PN M 1-20 we have no spatial information from the {\it Spitzer}
spectra (which covered the entire nebula) and the 3.3 $\mu$m emission in the
VLT/ISAAC spectra is not extended.

In short, still it is not clear if the emission bands at 3.3 and 11.3 $\mu$m
seen in PNe with fullerenes are only due to pure aromatic compounds or to
mixed aromatic/aliphatic structures such as those of HAC-like dust.


\section{Non detection of the fullerane features}

The IR laboratory spectra (R$\sim$500) of several fulleranes such as
C$_{60}$H$_{18}$, C$_{60}$H$_{36}$ or C$_{70}$H$_{38}$ (Iglesias-Groth et al.
2012), show that the strongest features in the mid-IR (2-20 $\mu$m) are those at
$\sim$3.44, 3.51 and 3.54 $\mu$m\footnote{Fulleranes do not emit at 3.3
$\mu$m  because they lack sp$^{2}$-bonded CH groups (see Figure 3).} (see Figure
3), being the best IR bands for searching these molecules in the circumstellar
environments of fullerene-rich PNe. Thus, at the resolution (R$\sim$600) of the
VLT/ISAAC observations it is possible to easily resolve these bands as well as
to distinguish them from the 3.3 $\mu$m emission (Iglesias-Groth et al. 2012).
However, none of these three emission features is detected in our VLT/ISAAC
spectra of Tc 1 and M 1-20. Figure 3 shows the non-detection of the
fullerane features between 3.4 and 3.6 $\mu$m in our VLT/ISAAC spectra of
Tc 1 and M 1-20. The laboratory spectra of the fulleranes
C$_{60}$H$_{18}$, C$_{70}$H$_{38}$ and C$_{60}$H$_{x}$+C$_{70}$H$_{y}$ (see
below) are also displayed and their corresponding 3.4-3.6 $\mu$m features
are marked.

We could estimate approximate upper limits to the fluxes of the hydrogenated fullerene
features (see Table 3). In order to obtain 2$\sigma$ upper limits to the expected
emission line fluxes of the fullerane features at $\sim$3.5 $\mu$m, we measure the
rms in our flux calibrated VLT/ISAAC spectra; rms values of $\sim$4.28 $\times$
10$^{-19}$ ergcm$^{-2}$s$^{-1}$\AA$^{-1}$ and $\sim$6.65 $\times$ 10$^{-18}$
ergcm$^{-2}$s$^{-1}$\AA$^{-1}$ for Tc 1 and M 1-20 are obtained, respectively. Then we
multiply these values by the widths (FWHMs in the range 0.02-0.10 $\mu$m) of the
fullerane bands measured in the infrared laboratory spectra\footnote{Note that
Zhang \& Kwok (2013) measured similar FWHMs (in the range $\sim$0.02-0.04 $\mu$m) for the
posssible fullerane bands seen in the ISO spectrum of IRAS 01005$+$7910.}. We have
used the laboratory spectra of two C$_{60}$H$_{18}$ isomers (obtained using hydrogen
iodide and direct hydrogenation with metal hydrides), two C$_{60}$H$_{36}$ spectra at
$+$48$^\circ$C and $+$250$^\circ$C, two C$_{70}$H$_{38}$ spectra at $+$50$^\circ$C and
$+$160$^\circ$C, and the fullerane mixture 77 \% of C$_{60}$H$_{x}$ and 22 \%
C$_{70}$H$_{y}$ with x $\approx$ y $\geq$ 30 at $+$45$^\circ$C (see Iglesias-Groth et al.
2012 for more details). Finally, we divided these fluxes by the area ($\sim$0.6"
arcsec$^{2}$ for both PNe) of the emission in the 3$-$4 $\mu$m range covered by our ISAAC
observations.

\begin{figure}[h]
   \centering
   \includegraphics[angle=0,scale=.50]{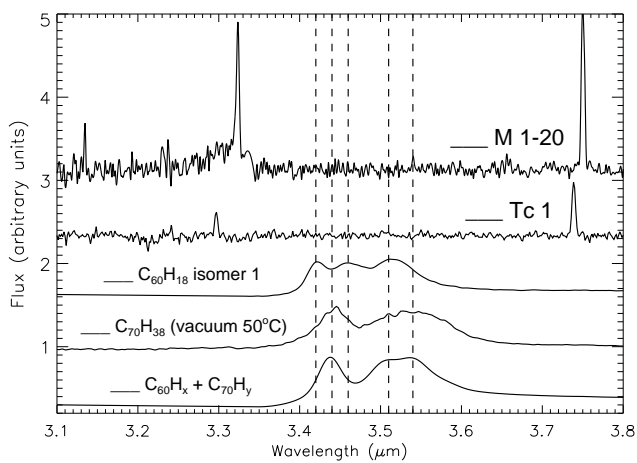}
    \caption{VLT/ISAAC spectra of M 1-20 and Tc 1 in comparison with the
    laboratory spectra of the fulleranes C$_{60}$H$_{18}$,
    C$_{70}$H$_{38}$, and C$_{60}$H$_{x}$+C$_{70}$H$_{y}$ in the 3.1-3.8 $\mu$m
    range. Note that the C$_{60}$H$_{18}$ fullerane bands at 3.42, 3.46,
    and 3.51 $\mu$m and the C$_{70}$H$_{38}$ fullerane bands at 3.44,
    and 3.54 $\mu$m are marked with  dashed lines. The
    C$_{60}$H$_{x}$+C$_{70}$H$_{y}$ fullerane bands are those at 3.44,
    3.51,  and 3.54 $\mu$m.\label{Fig3}}
   \end{figure}
   
\begin{table*}
\tiny
\caption{\label{t3}Summary table showing examples of the expected fluxes for the fullerane features. Quoted values are the measured 
central wavelengths of the fullerane features, the full widths at half maximum (FWHM), the
2$\sigma$ upper detection limits to the expected emission line fluxes in Tc 1 and M 1-20, and
the predicted fluxes for the fullerane bands.}
\centering
\small\begin{tabular}{lcccccc}
\hline\hline
 Fullerane & $\lambda$$_{c}$ & FWHM              & FLUX\tablefootmark{a} (Tc 1)      & Predicted fluxes (Tc 1)          & FLUX\tablefootmark{a} (M 1-20)   &   Predicted fluxes (M 1-20)       \\
           &  ($\mu$m)       & (10$^{-2}$$\mu$m) & (10$^{-16}$ergcm$^{-2}$s$^{-1}$/"$^{2}$)  & (10$^{-14}$ergcm$^{-2}$s$^{-1}$/"$^{2}$) & (10$^{-15}$ergcm$^{-2}$s$^{-1}$/"$^{2}$) & (10$^{-13}$ergcm$^{-2}$s$^{-1}$/"$^{2}$) \\
\hline
  C$_{60}$H$_{18}$ (isomer 1)                    & 3.42 & 2.1  & $\leq$3.00  &     9.43\tablefootmark{b}  & $\leq$4.65  &  2.39\tablefootmark{b} \\
                                                 & 3.46 & 3.6  & $\leq$5.13  &    10.20\tablefootmark{b}  & $\leq$7.98  &  2.58\tablefootmark{b} \\
                                                 & 3.51 & 5.1  & $\leq$7.28  &    13.25\tablefootmark{b}  & $\leq$11.30 &  3.36\tablefootmark{b} \\   
  C$_{60}$H$_{18}$ (isomer 2)                    & 3.42 & 2.2  & $\leq$3.13  &    12.74\tablefootmark{c}  & $\leq$4.87  &  1.42\tablefootmark{c} \\
                                                 & 3.45 & 3.4  & $\leq$4.85  &    13.92\tablefootmark{c}  & $\leq$7.53  &  1.56\tablefootmark{c} \\
                                                 & 3.51 & 5.5  & $\leq$7.85  &    18.21\tablefootmark{c}  & $\leq$12.18 &  2.06\tablefootmark{c} \\
  C$_{60}$H$_{36}$ (+48$^\circ$C)                & 3.44 & 2.3  & $\leq$3.33  &    22.13\tablefootmark{b}  & $\leq$5.18  &  5.58\tablefootmark{b} \\	
                                                 & 3.54 & 7.7  & $\leq$11.02 &    16.23\tablefootmark{b}  & $\leq$17.10 &  4.11\tablefootmark{b} \\    
  C$_{60}$H$_{36}$ (+250$^\circ$C)               & 3.44 & 2.2  & $\leq$3.10  &    30.11\tablefootmark{c}  & $\leq$4.80  &  3.39\tablefootmark{c} \\ 
                                                 & 3.53 & 10.1 & $\leq$14.38 &    22.08\tablefootmark{c}  & $\leq$22.33 &  2.47\tablefootmark{c} \\
  C$_{70}$H$_{38}$ (+50$^\circ$C)                & 3.44 & 2.5  & $\leq$3.62  &    18.13\tablefootmark{b}  & $\leq$5.63  &  4.61\tablefootmark{b} \\
                                                 & 3.54 & 7.0  & $\leq$10.05 &    16.75\tablefootmark{b}  & $\leq$15.62 &  4.22\tablefootmark{b} \\ 
  C$_{70}$H$_{38}$ (+160$^\circ$C)               & 3.44 & 2.5  & $\leq$3.55  &     1.81\tablefootmark{d}  & $\leq$5.52  &  2.72\tablefootmark{c} \\ 
                                                 & 3.54 & 6.3  & $\leq$8.97  &     1.68\tablefootmark{d}  & $\leq$13.93 &  2.56\tablefootmark{c} \\
  C$_{60}$H$_{x}$+C$_{70}$H$_{y}$ (+45$^\circ$C) & 3.44 & 2.2  & $\leq$3.13  &    20.13\tablefootmark{b}  & $\leq$4.87  &  5.08\tablefootmark{b} \\
                                                 & 3.51 & 2.8  & $\leq$4.00  &			          & $\leq$6.20  &   		      \\
                                                 & 3.54 & 2.5  & $\leq$3.57  &    18.71\tablefootmark{b}  & $\leq$5.53  &  4.75\tablefootmark{b} \\
\hline\hline
\end{tabular}
\tablefoot{
\\
\tablefoottext{a}{Estimated flux errors are $\sim$30\%-40\%.}
\\
\tablefoottext{b}{Flux prediction obtained by taking the observed flux of the C$_{60}$ feature
at $\sim$8.5 $\mu$m.}
\\
\tablefoottext{c}{Flux prediction obtained by taking the observed flux of the C$_{60}$ feature
at $\sim$17.4 $\mu$m.}
\\
\tablefoottext{d}{Flux prediction obtained by taking the observed flux of the C$_{70}$ feature
at $\sim$14.9 $\mu$m.}
}
\end{table*}

On the other hand, by using the molar absorptivity values for the fullerenes and
fulleranes reported by Iglesias-Groth et al. (2011, 2012), we may
estimate the predicted fluxes for the fullerane features in Tc 1 and M
1-20\footnote{The molar absorptivities of fullerenes and fulleranes are
determined with the same standard procedure (Iglesias-Groth et al. 2011, 2012)
and we have used them to estimate the predicted fluxes.} (see Table 3). By
taking the molar absorptivity ratio (e.g.,
$\epsilon$$_{C60}$/$\epsilon$$_{fulleranes}$) of the fullerene and
fullerane bands and the observed fluxes of the C$_{60}$ and C$_{70}$
infrared bands less contaminated by other species (Garc\'{\i}a-Hern\'andez et
al. 2012) we estimate the expected flux of the fullerane features at
$\sim$3.5 $\mu$m. It is to be noted here that this is only an approximation
and a complete model for the excitation/emission of both species should be
needed. The {\it Spitzer} integrated fluxes of the infrared bands at 8.5 and
17.4 $\mu$m (C$_{60}$), and 14.9 $\mu$m (C$_{70}$) were taken from
Garc\'{\i}a-Hern\'andez et al. (2012) and converted to flux/arcsec$^{2}$ by
considering the area covered by the fullerenes emission in the {\it Spitzer}
spectra of  Tc 1 ($\sim$80 arcsec$^{2}$ at 8.5 $\mu$m and $\sim$53 arcsec$^{2}$
at 14.9 and 17.4 $\mu$m) and M 1-20 ($\sim$4 arcsec$^{2}$ at 8.5, 14.9, and 17.4
$\mu$m). 

Table 3 shows some examples of the estimated 2$\sigma$ upper limits and the predicted
fluxes for the fullerane bands. For example, the molar absorptivity ratio of the
C$_{60}$H$_{36}$ band at 3.44 $\mu$m and the C$_{60}$ infrared band at 8.5 $\mu$m is
$\epsilon$$_{8.5}$/$\epsilon$$_{3.44}$$\sim$0.34. Thus, the  expected flux of the
C$_{60}$H$_{36}$ band at 3.44 $\mu$m in Tc 1 and M 1-20 would be $\sim$22.13 $\times$
10$^{-14}$ ergcm$^{-2}$s$^{-1}$/arcsec$^{2}$ and $\sim$5.58 $\times$ 10$^{-13}$
ergcm$^{-2}$s$^{-1}$/arcsec$^{2}$, respectively (see Table 3). Similar values are
obtained for the fullerane C$_{60}$H$_{36}$ band at 3.54 $\mu$m, with expected
fluxes of $\sim$16.23 $\times$ 10$^{-14}$ ergcm$^{-2}$s$^{-1}$/arcsec$^{2}$, and
$\sim$4.11 $\times$ 10$^{-13}$ ergcm$^{-2}$s$^{-1}$/arcsec$^{2}$ in Tc 1 and M 1-20,
respectively. In summary, the expected fluxes for all fullerane bands of
C$_{60}$H$_{18}$, C$_{60}$H$_{36}$, C$_{70}$H$_{38}$ and the mixture
C$_{60}$H$_{x}$+C$_{70}$H$_{y}$ (by using the observed {\it Spitzer} fluxes of the
C$_{60}$ 8.5 and 17.4 $\mu$m and C$_{70}$ 14.9 $\mu$m bands) are in the range of
$\sim$1.7-30 $\times$ 10$^{-14}$ and $\sim$1.4-5.6 $\times$ 10$^{-13}$
ergcm$^{-2}$s$^{-1}$/arcsec$^{2}$ in Tc 1 and M 1-20, respectively (see Table 3 for some
representative examples).

By comparing the values of the predicted fluxes with our 2$\sigma$ upper limits (Table
3), we find that the expected fluxes are a factor of $\sim$20$-$1000 and $\sim$10$-$100
higher than the 2$\sigma$ upper limits for Tc 1 and M 1-20, respectively. From these
estimations we thus conclude that if fulleranes are present in Tc 1 and M 1-20,
then they seem to be by far less abundant than C$_{60}$ and C$_{70}$.

As we mentioned above, thermal heating via chemical reactions internal to HAC dust may
explain the detection of fullerenes in the H-rich circumstellar environments of PNe and
would potentially form fulleranes (Duley \& Williams 2011). In addition, Duley \&
Hu (2012) reported that laboratory spectra of HAC nano-particles containing fullerene
precursors (or proto-fullerenes; PFs), but not C$_{60}$, display the same set of mid-IR
features as the isolated C$_{60}$ molecule. They suggest an evolutionary sequence for the
conversion of HAC to fullerenes, in which initial HAC de-hydrogenation is followed by PFs
formation and subsequent conversion of PFs to closed cage structures such as C$_{60}$.
Under the Duley \& Hu (2012) scenario, Tc 1 would represent the last stage in the HAC
de-hydrogenation process (i.e., only fullerenes are present), while the proto-PN IRAS
01005$+$7910 would represent an intermediate stage where the PFs are being converted to
fullerenes. Remarkably, fulleranes may be also by-products of this conversion from
HAC to fullerenes (Duley \& Hu 2012).

Interestingly, Zhang \& Kwok (2013) have (tentatively) detected fulleranes in the
proto-PN IRAS 01005$+$7910; three strong C-H stretching bands at 3.48, 3.51, and 3.58
$\mu$m are apparently present in its ISO spectrum with fluxes comparable to the one of
the 3.3 $\mu$m feature. By assuming that all features have the same oscillator strength,
they conclude that the calculated relative strengths indicate that the m value (degree of
hydrogenation of the fulleranes C$_{60}$H$_{m}$) may lie within the range from
25-40, which is consistent with the production of C$_{60}$H$_{36}$ (the dominant product
of the hydrogenation reaction of C$_{60}$; e.g., Cataldo \& Iglesias-Groth
2009)\footnote{Zhang \& Kwok (2013) suggest that C$_{60}$H$_{18}$ might be also
present.}. In addition, the observed fluxes of the C$_{60}$ and fullerane bands
suggest that about 50 percent of fullerenes have been hydrogenated. Curiously, IRAS
01005$+$7910 exceptionally displays very strong C$_{60}$$^+$ DIBs (Iglesias-Groth \&
Esposito 2013).

Our non-detection of fulleranes in more evolved PNe (such as Tc 1 and M
1-20 with T$_{eff}$$>$ 30,000 K) together with their possible detection in
less evolved sources (like IRAS 01005$+$7910 with T$_{eff}$$\sim$21,500 K) thus
suggest that fulleranes may be formed in the short transition phase
between AGB stars and PNe but they are quickly destroyed; e.g., photochemically
processed by the rapidly changing UV radiation from the central star. The rapid
increase of the UV radiation towards the PN stage will easily break the C$-$H
bonds of the previously formed fullerane molecules, forming
fulleranes at lower hydrogenation degree or forming back - more resistant
- C$_{60}$ and C$_{70}$ fullerenes and molecular hydrogen (Cataldo \&
Iglesias-Groth 2009). However, still it is not clear what is the origin of
fulleranes. Fulleranes could be formed via the Duley \& Hu (2012)
HAC de-hydrogenation process mentioned above in which fulleranes may be
by-products of the conversion from PFs to C$_{60}$ or they could be formed by
the reaction of pre-existing fullerenes (e.g., formed by the photochemical
processing of HACs; Scott et al. 1997) with atomic hydrogen. During the short
($\sim$10$^{2}$-10$^{4}$ years) transition phase AGB-PN, H$_{2}$\footnote{The
HAC chemical energy model of Duley \& Williams (2011) predict also the release
of warm H$_{2}$ molecules trapped inside the HAC solid.} may be photodissociated
into atomic H due to the gradual increase of the UV radiation field with the
evolution of the central star and/or by shocks (H$_{2}$ dissociated through
collisions) due to the fast post-AGB stellar winds. By reacting with atomic
hydrogen fullerenes form fulleranes at various degree of hydrogenation.
Thus, the post-AGB phase would make the reaction between fullerenes and atomic H
more favorable (Cataldo \& Iglesias-Groth 2009) and it seems to provide the
right conditions for fullerane formation and detection.


\section{Conclusions}

We have presented VLT/ISAAC 2.9-4.1 $\mu$m spectroscopy of two fullerene PNe in
our Galaxy. The spectrum of Tc 1 shows a continuum with no signs of UIE features
and only H (and He) nebular lines. This indicates that the IR bands seen
in the Tc 1 {\it Spitzer} spectrum are very likely due to C$_{60}$ and C$_{70}$
solely, and not other kind of molecules such as fulleranes or
fullerene/PAH adducts. The M 1-20 VLT/ISAAC spectrum shows H and He nebular
lines in conjunction with the UIE feature at 3.3 $\mu$m.

The VLT/ISAAC and {\it Spitzer} spectra of these fullerene PNe confirm a
correlation between the 3.3 and 11.3 $\mu$m UIE features previously reported in
the literature, which suggests a common carrier for these two features.
Unfortunately, the VLT/ISAAC observations presented here cannot reveal the
nature of the real carrier (e.g., PAHs vs. HACs) of the 3.3 $\mu$m emission. 

Interestingly, we have reported the non-detection of the strongest bands of hydrogenated
fullerenes (fulleranes such as C$_{60}$H$_{18}$, C$_{60}$H$_{36}$,
C$_{70}$H$_{38}$ and a fullerane mixture) at $\sim$3.44, 3.51 and 3.54 $\mu$m in
the 3$-$4 $\mu$m spectra of the fullerene-containing PNe Tc 1 and M 1-20. From the
comparison of the predicted fluxes of the fullerane bands with our 2$\sigma$ upper
limits, we conclude that if fulleranes are present in both objects, then they seem
to be by far less abundant than isolated fullerene molecules. 

Our non-detection of fulleranes in two fullerene PNe together with their
possible detection (if real) in the fullerene proto-PN IRAS 01005$+$7910 suggest
that these fullerene-related species may be formed in the short transition phase
AGB$-$PN but they are rapidly destroyed; e.g., by the quick increase of the UV
radiation from the central star towards the PN stage. The transition between AGB
stars and PNe seems to be the best evolutionary stellar phase for finding
fulleranes in space and 3$-$4 $\mu$m spectroscopy in a larger sample of
C-rich proto-PNe is encouraged.


\begin{acknowledgements}

We acknowledge Kameswara Rao for his help during the data analysis. J.J.D.L.,
D.A.G.H., and A.M. acknowledge support provided by the Spanish Ministry of
Economy and Competitiveness (MINECO) under grant AYA$-$2014$-$58082$-$P. D.A.G.H.
was funded by the Ram\'on y Cajal fellowship number RYC$-$2013$-$14182. This work
is based on observations obtained with ESO/VLT under the programme 290.D-5093(A). 

\end{acknowledgements}



\begin{thebibliography}{22}
\expandafter\ifx\csname natexlab\endcsname\relax\def\natexlab#1{#1}\fi

\bibitem[{{Abia} {et~al.}(2001){Abia}, {Busso}, {Gallino}, {Dom{\'{\i}}nguez},
  {Straniero}, \& {Isern}}]{abia2001}
{Abia}, C., {Busso}, M., {Gallino}, R., {et~al.} 2001, \apj, 559, 1117

\bibitem[{{Alvarez} \& {Plez}(1998)}]{alvarez1998}
{Alvarez}, R. \& {Plez}, B. 1998, \aap, 330, 1109

\bibitem[{{Busso} {et~al.}(1999){Busso}, {Gallino}, \&
  {Wasserburg}}]{busso1999}
{Busso}, M., {Gallino}, R., \& {Wasserburg}, G.~J. 1999, \araa, 37, 239

\bibitem[{{Decin} {et~al.}(2010){Decin}, {Justtanont}, {De Beck}, {Lombaert},
  {de Koter}, {Waters}, {Marston}, {Teyssier}, {Sch{\"o}ier}, {Bujarrabal},
  {Alcolea}, {Cernicharo}, {Dominik}, {Melnick}, {Menten}, {Neufeld},
  {Olofsson}, {Planesas}, {Schmidt}, {Szczerba}, {de Graauw}, {Helmich},
  {Roelfsema}, {Dieleman}, {Morris}, {Gallego}, {D{\'{\i}}ez-Gonz{\'a}lez}, \&
  {Caux}}]{decin2010}
{Decin}, L., {Justtanont}, K., {De Beck}, E., {et~al.} 2010, \aap, 521, L4

\bibitem[{{Garc{\'{\i}}a-Hern{\'a}ndez}
  {et~al.}(2006){Garc{\'{\i}}a-Hern{\'a}ndez}, {Garc{\'{\i}}a-Lario}, {Plez},
  {D'Antona}, {Manchado}, \& {Trigo-Rodr{\'{\i}}guez}}]{garcia2006}
{Garc{\'{\i}}a-Hern{\'a}ndez}, D.~A., {Garc{\'{\i}}a-Lario}, P., {Plez}, B.,
  {et~al.} 2006, Science, 314, 1751

\bibitem[{{Garc{\'{\i}}a-Hern{\'a}ndez}
  {et~al.}(2007){Garc{\'{\i}}a-Hern{\'a}ndez}, {Garc{\'{\i}}a-Lario}, {Plez},
  {Manchado}, {D'Antona}, {Lub}, \& {Habing}}]{garcia2007}
{Garc{\'{\i}}a-Hern{\'a}ndez}, D.~A., {Garc{\'{\i}}a-Lario}, P., {Plez}, B.,
  {et~al.} 2007, \aap, 462, 711

\bibitem[{{Garc{\'{\i}}a-Hern{\'a}ndez}
  {et~al.}(2009){Garc{\'{\i}}a-Hern{\'a}ndez}, {Manchado}, {Lambert}, {Plez},
  {Garc{\'{\i}}a-Lario}, {D'Antona}, {Lugaro}, {Karakas}, \& {van
  Raai}}]{garcia2009}
{Garc{\'{\i}}a-Hern{\'a}ndez}, D.~A., {Manchado}, A., {Lambert}, D.~L.,
  {et~al.} 2009, \apjl, 705, L31

\bibitem[{{Gray}(2008)}]{gray2008}
{Gray}, D.~F. 2008, {The Observation and Analysis of Stellar Photospheres}

\bibitem[{{Grevesse} {et~al.}(2007){Grevesse}, {Asplund}, \&
  {Sauval}}]{grevesse2007}
{Grevesse}, N., {Asplund}, M., \& {Sauval}, A.~J. 2007, \ssr, 130, 105

\bibitem[{{Grevesse} \& {Sauval}(1998)}]{grevesse1998}
{Grevesse}, N. \& {Sauval}, A.~J. 1998, Space Science Reviews, 85, 161

\bibitem[{{Justtanont} {et~al.}(2013){Justtanont}, {Teyssier}, {Barlow},
  {Matsuura}, {Swinyard}, {Waters}, \& {Yates}}]{justtanont2013}
{Justtanont}, K., {Teyssier}, D., {Barlow}, M.~J., {et~al.} 2013, \aap, 556,
  A101

\bibitem[{{Karakas} {et~al.}(2012){Karakas}, {Garc{\'{\i}}a-Hern{\'a}ndez}, \&
  {Lugaro}}]{karakas2012}
{Karakas}, A.~I., {Garc{\'{\i}}a-Hern{\'a}ndez}, D.~A., \& {Lugaro}, M. 2012,
  \apj, 751, 8

\bibitem[{{Lambert} {et~al.}(1995){Lambert}, {Smith}, {Busso}, {Gallino}, \&
  {Straniero}}]{lambert1995}
{Lambert}, D.~L., {Smith}, V.~V., {Busso}, M., {Gallino}, R., \& {Straniero},
  O. 1995, \apj, 450, 302

\bibitem[{{Mihalas} \& {Kunasz}(1986)}]{mihalas1986}
{Mihalas}, D. \& {Kunasz}, P.~B. 1986, Journal of Computational Physics, 64, 1

\bibitem[{{Plez} {et~al.}(1992){Plez}, {Brett}, \& {Nordlund}}]{plez1992}
{Plez}, B., {Brett}, J.~M., \& {Nordlund}, A. 1992, \aap, 256, 551

\bibitem[{{Plez} \& {Lambert}(2002)}]{plez2002}
{Plez}, B. \& {Lambert}, D.~L. 2002, \aap, 386, 1009

\bibitem[{{Plez} {et~al.}(1993){Plez}, {Smith}, \& {Lambert}}]{plez1993}
{Plez}, B., {Smith}, V.~V., \& {Lambert}, D.~L. 1993, \apj, 418, 812

\bibitem[{{Sackmann} \& {Boothroyd}(1992)}]{sackmann1992}
{Sackmann}, I.-J. \& {Boothroyd}, A.~I. 1992, \apjl, 392, L71

\bibitem[{{Schwarzschild} \& {H{\"a}rm}(1967)}]{sch1967}
{Schwarzschild}, M. \& {H{\"a}rm}, R. 1967, \apj, 150, 961

\bibitem[{{Tomkin} \& {Lambert}(1999)}]{tomkin1999}
{Tomkin}, J. \& {Lambert}, D.~L. 1999, \apj, 523, 234

\bibitem[{{Truran} \& {Iben}(1977)}]{truran1977}
{Truran}, J.~W. \& {Iben}, Jr., I. 1977, \apj, 216, 797

\bibitem[{{van Raai} {et~al.}(2012){van Raai}, {Lugaro}, {Karakas},
  {Garc{\'{\i}}a-Hern{\'a}ndez}, \& {Yong}}]{vanraai2012}
{van Raai}, M.~A., {Lugaro}, M., {Karakas}, A.~I.,
  {Garc{\'{\i}}a-Hern{\'a}ndez}, D.~A., \& {Yong}, D. 2012, \aap, 540, A44

\end{thebibliography}


\begin{thebibliography}{}

\bibitem[]{allamandola} Allamandola, L. J., Tielens, A. G. G. M., \& Barker, J. R. 1985, ApJ, 290, L25
\bibitem[]{bauschlicher} Bauschlicher, C. W., Boersma, C., Ricca, A. et al. 2010, ApJS, 189, 341
\bibitem[]{beaulieu} Beaulieu, S. F., Dopita, M. A., \& Freeman, K. C. 1999, ApJ, 515, 610
\bibitem[]{bernardsalas} Bernard-Salas, J., Cami, J., Peeters, E. et al. 2012, ApJ, 757, 41
\bibitem[]{boersma} Boersma, C., Bauschlicher, C. W., Ricca, A. et al. 2014, ApJS, 211, 8
\bibitem[]{cami10} Cami, J. et al. 2010, Science, 329, 1180
\bibitem[]{cataldo} Cataldo, F., \& Iglesias-Groth, S. 2009, MNRAS, 400, 291 
\bibitem[]{cataldo} Cataldo, F., Garc\'{\i}a-Hern\'andez, D. A., \& Manchado, A. 2014, Eur. Chem. Bull., 3, 740
\bibitem[]{cataldo} Cataldo, F., Garc\'{\i}a-Hern\'andez, D. A., \& Manchado, A. 2015, FNCN, 23, 818-823
\bibitem[]{chiar} Chiar, J. E., Tielens, A. G. G. M., Whittet, D. C. B. et al. 2000, ApJ, 537, 749
\bibitem[]{cox} Cox, N. L. J. 2011, in PAHs and the Universe: A Symposium to Celebrate the 25th Anniversary of the PAH Hypothesis, EAS Publication Series Vol. 46, 349-354, eds Joblin, C. \& Tielens, A. G. G. M.
\bibitem[]{cutri} Cutri, R. M. et al. 2012, VizieR On-line Data Catalog: II/311
\bibitem[]{debruijne} De Bruijne, J. H. J., \& Eilers, A. -C. 2012, A\&A, 546, A61
\bibitem[]{diaz} D\'{\i}az-Luis, J. J., Garc\'{\i}a-Hern\'andez, D. A., D. A., Rao, N. K. et al. 2015, A\&A, 573, A97
\bibitem[]{duley} Duley, W. W., \& Williams, D. A. 1981, MNRAS, 196, 269
\bibitem[]{duley} Duley, W. W., \& Williams, D. A. 2011, ApJ, 737, L44
\bibitem[]{duley} Duley, W. W., \& Hu, A. 2012, ApJ, 745, L11
\bibitem[]{gadallah} Gadallah, K. A. K., Mutschke, H., \& Jager, C. 2013, A\&A, 554, A12
\bibitem[]{garcia1} Garc\'{\i}a-Hern\'andez, D. A., Manchado, A., Garc\'{\i}a-Lario, P. et al. 2010, ApJ, 724, L39 
\bibitem[]{garcia4} Garc\'{\i}a-Hern\'andez, D. A., Iglesias-Groth, S., Acosta-Pulido, J. A. et al. 2011a, ApJ, 737, L30
\bibitem[]{garcia2} Garc\'{\i}a-Hern\'andez, D. A., Rao, N. K., \& Lambert, D. L. 2011b, ApJ, 729, 126
\bibitem[]{garcia6} Garc\'{\i}a-Hern\'andez, D. A., Villaver, E., Garc\'{\i}a-Lario, P. et al. 2012, ApJ, 760, 107
\bibitem[]{garcia8} Garc\'{\i}a-Hern\'andez, D. A., Cataldo, F., \& Manchado, A. 2013, MNRAS, 434, 415
\bibitem[]{garcia7} Garc\'{\i}a-Hern\'andez, D. A., \& D\'{\i}az-Luis, J. J. 2013, A\&A, 550, L6
\bibitem[]{garcia} Garc\'{\i}a-Lario, P., Manchado, A., Ulla, A., Manteiga, M. 1999, ApJ, 513, 941
\bibitem[]{gudennavar} Gudennavar, S. B., Bubbly, S. G., Preethi, K. et al. 2012, ApJ, 199, 8
\bibitem[]{iglesias} Iglesias-Groth, S. 2007, ApJ, 661, L167
\bibitem[]{iglesias} Iglesias-Groth, S., Cataldo, F., \& Manchado, A. 2011, MNRAS, 413, 213
\bibitem[]{iglesias} Iglesias-Groth, S., Garc\'{\i}a-Hern\'andez, D. A., Cataldo, F. et al. 2012, MNRAS, 423, 2868
\bibitem[]{iglesias} Iglesias-Groth, S., \& Esposito, M. 2013, ApJ, 776, L2
\bibitem[]{jourdaindemuizon} Jourdain de Muizon, M., D'Hendecourt, L. B., \& Geballe, T. R. 1990, A\&A, 227, 526
\bibitem[]{kroto85} Kroto, H. W., Heath, J. R., Obrien, S. C. et al. 1985, Nature, 318, 162
\bibitem[]{kwok} Kwok, S., Volk, K., \& Hrivnak, B. J. 1989, ApJ, 345, L51
\bibitem[]{kwok} Kwok, S., Volk, K., \& Hrivnak, B. J. 1999, A\&A, 350, L35
\bibitem[]{kwok} Kwok, S., \& Zhang, Y. 2011, Nature, 479, 80
\bibitem[]{leger} Leger, A., \& Puget, J. L. 1984, A\&A, 137, L5
\bibitem[]{maltseva} Maltseva, E., Petrignani, A., Candian, A. et al. 2015, eprint arXiv:1510.04948
\bibitem[]{otsuka} Otsuka, M., Kemper, F., Cami, J. et al. 2014, MNRAS, 437, 2577
\bibitem[]{russell} Russell, R. W., Soifer, B. T., \& Merrill, K. M. 1977, ApJ, 213, 66
\bibitem[]{sadjadi} Sadjadi, S. A., Zhang, Y., \& Kwok, S. 2015, ApJ, 807, 95
\bibitem[]{scott} Scott, A., Duley, W. W., \& Pinho, G. P. 1997, ApJ, 489, L193
\bibitem[]{sellgren} Sellgren, K., Uchida, K. I., \& Werner, M. W. 2007, ApJ, 659, 1338
\bibitem[]{sellgren} Sellgren, K., Werner, M. W., Ingalls, J. G. et al. 2010, ApJ, 722, L54
\bibitem[]{smith} Smith, E. C. D., \& McLean, I. S. 2008, ApJ, 676, 408
\bibitem[]{soubiran} Soubiran, C., Jasniewicz, G., Chemin, L. et al. 2013, A\&A, 552, A64
\bibitem[]{sturm} Sturm, E., Lutz, D., Tran, D. et al. 2000, A\&A, 358, 481
\bibitem[]{tokunaga} Tokunaga, A. T., Sellgren, K., Smith, R. G. et al. 1991, ApJ, 380, 452
\bibitem[]{volk} Volk, K., Hrivnak, B. J., Matsuura, M. et al. 2011, ApJ, 735, 127 
\bibitem[]{wang} Wang, W., \& Liu, X. -W. 2007, MNRAS, 381, 669
\bibitem[]{webster} Webster, A. 1995, MNRAS, 277, 1555
\bibitem[]{wegner} Wegner, W. 2003, AN, 324, 219
\bibitem[]{williams} Williams, R., Jenkins, E. B., Baldwin, J. A. et al. 2008, ApJ, 677, 1100 
\bibitem[]{zhang} Zhang, Y., \& Kwok, S. 2011, ApJ, 730, 126 
\bibitem[]{zhang} Zhang, Y., \& Kwok, S. 2013, Earth, Planets and Space, 65, 1069
 
\end{thebibliography}
\end{document}